\documentclass[prb,twocolumn]{revtex4-1}

\usepackage{amsmath}  
\usepackage{amsfonts} 
\usepackage{graphicx} 
\usepackage{siunitx}
\usepackage[colorlinks=true,
            linkcolor=blue,
            citecolor=blue,
            urlcolor=blue]{hyperref}
\usepackage[]{pdfpages}

\makeatletter
\AtBeginDocument{\let\LS@rot\@undefined}
\makeatother

\begin{document}


\title[Diffusion and disorder-induced localization]{Simulating diffusion and disorder-induced localization in random walks and transmission lines}

\author{Jake S.\ Bobowski}
\email{jake.bobowski@ubc.ca} 
\affiliation{Department of Physics, University of British Columbia, 3333 University Way, Kelowna, British Columbia, V1V 1V7, Canada}



\date{\today}

\begin{abstract}
We present two complementary simulations that lead to an exploration of Anderson localization, a phenomenon in which wave diffusion is suppressed in disordered media by interference from multiple scattering. To build intuition, the first models the random walk of classical, non-interacting point-like particles, providing a clear analogy to the way disorder can limit transport. The second examines the propagation of an electromagnetic pulse through a one-dimensional, lossless transmission line with randomly varying propagation constant and characteristic impedance along its length, a system that captures the interference effects essential for true Anderson localization. We evaluate quantitative measures that reveal the transition from normal diffusion to localization of particles in one case, and the exponential confinement of wave energy in the other.  Together, these simulations offer a pair of accessible tools for investigating localization phenomena in an instructional setting.
\end{abstract}


\maketitle 

\section{Introduction}\label{sec:intro} 

Disorder-induced confinement of electron wavefunctions, a phenomenon termed Anderson localization (AL), was first proposed by Philip Anderson in a landmark 1958 paper entitled {\it Absence of Diffusion in Certain Random Lattices}.~\cite{Anderson:1958}  In the presence of a random potential, the net amplitude for scattering an electron from point $a$ to $b$ is exponentially suppressed.  The intuitive explanation is that the potential randomizes the phase of each scattering path between the two points, which leads to nearly complete destructive interference when the individual scattering amplitudes are summed.  

AL provides one possible mechanism for metal-to-insulator transitions (MIT) in materials, such as doped semiconductors, in which an abrupt drop in conductivity, due to localization of charge carriers, is observed as temperature or doping is tuned through a critical value.~\cite{Yamanouchi:1967, Mott:1968}  It should be noted, however, that the current consensus is that both disorder-induced Anderson localization and interaction-driven Mott physics are required for a complete description of the MIT in doped semiconductors.\cite{Lee:1985}  Remarkable demonstrations of AL in quantum mechanical matter waves were achieved in two separate 2008 studies of Bose-Einstein condensates (BECs).  In one case, spatial confinement of the BEC released into a one-dimensional (1D) waveguide was directly observed.\cite{Billy:2008} In the second case, in the presence of sufficient disorder, a BEC of non-interacting atoms on a 1D lattice was observed to maintain its size after release from its trap.\cite{Roati:2008}   Not long after, it was realized that, although AL was originally formulated in terms of quantum mechanical electron wavefunctions, the same reasoning can likewise be applied to classical waves in disordered media (electromagnetic or acoustic).\cite{Chabanov:2000, Hu:2008, Lagendijk:2009}.  

In this paper, rather than re-present a full theoretical treatment of AL, we briefly review only the transmission line (TL) theory needed to support our simulations.  For comprehensive discussions of AL, readers are referred to a number of high-quality review articles on the subject.  References [\onlinecite{Lagendijk:2009}] and [\onlinecite{Mafi:2015}] provide pedagogical overviews, while Refs.~[\onlinecite{Lee:1985}] and [\onlinecite{Wolfle:2010}] offer more technical treatments. 

\section{Anderson localization and Pedagogy}
It is important to emphasize that AL requires neither interactions between particles nor a loss mechanism: it is an emergent effect arising solely from disorder, with scattering that can be entirely energy-conserving. The study of emergent collective phenomena is pervasive in modern physics research, particularly in condensed matter physics.\cite{Anderson:1972, Kivelson:2016} By contrast, undergraduate curricula, with a few exceptions such as BEC, tend to emphasize single-particle physics or weakly coupled systems.  Our primary motivation for this work is to provide an accessible means to discuss AL as another example of emergent behavior that is not an obvious extension of the underlying fundamental laws governing the individual constituents of the system.  While the landmark studies discussed in Sec.~\ref{sec:intro} firmly established AL across condensed matter, ultracold atomic, and classical wave systems, the essential ideas can be difficult for students to access without specialized equipment and, in many cases, without the mathematical background to fully engage with the theory.  Numerical modeling provides an attractive alternative: carefully chosen simulations can reveal the underlying mechanisms of localization in ways that are both intuitive and quantitative. In this work, we present two complementary approaches. The first illustrates how disorder suppresses diffusion through a random walk of classical particles, while the second, which is our primary focus, examines electromagnetic pulse propagation along 1D transmission lines with random impedance and phase velocity, capturing the multiple-scattering interference central to true Anderson localization. Together, these models provide flexible and accessible tools for exploring localization phenomena in an instructional setting.

A number of previous contributions to the {\it American Journal of Physics} (AJP) have explored Anderson localization from a pedagogical perspective. Early efforts demonstrated localization of vibrational wave packets in disordered chains and introduced simplified tight-binding models to highlight scaling and localization length in 1D systems.\cite{Allen:1998, Dominguez-Adame:2004, Siber:2006} Transmission line analogies have also been employed to illustrate the emergence of band gaps and localization in periodic, quasiperiodic, and random lattices.\cite{Gutierrez-Medina:2013} More recently, experimental approaches have been developed for the undergraduate laboratory, including demonstrations of disorder-induced suppression of wave transport in stacks of glass slides.\cite{Kemp:2016} Finally, accessible treatments of disorder in solid-state physics have been proposed using transfer matrix methods and the coherent potential approximation.\cite{Martinez:2023}

Our work differs in two respects. First, we provide a two-level pedagogical approach: beginning with a point-particle random walk model that connects directly to undergraduate treatments of diffusion and Brownian motion, and then advancing to a full transmission line simulation that explicitly captures the multiple-scattering interference central to AL. Second, our simulations emphasize time-domain pulse propagation and energy confinement, providing visualizations (heatmaps, ensemble averages) that go beyond static transmission coefficients. Together, these features complement and extend the existing AJP literature, offering an accessible entry point for instructors and students alike.  Having established the pedagogical context, we next introduce our first simulation: a random walk model of diffusion suppression.

\section{Disorder-Suppressed Diffusion}
In our physics major, students complete a second-year Modern Physics Laboratory course which has a lecture component that covers the fundamentals of data analysis.  In the lecture, we discuss the binomial distribution and 1D random walk which provides (1) a path to the Poisson and Gaussian distributions and (2) the theoretical framework for a Brownian motion lab in which students study the dynamics of micron-sized latex spheres suspended in a saline solution.  We initially developed a simple non-interacting point particle simulation using the Python programming language to demonstrate diffusion.  We placed $10^4$ particles at the center of a $1\times 1$ square domain and allowed them to undergo an outward random walk from that point, analogous to placing a drop of food coloring in the center of a cup of water.  Each iteration of the simulation assigned particle $i$ a Gaussian-distributed step $\Delta \mathbf{r}_i = \Delta x_i\,\hat\imath + \Delta y_i\,\hat\jmath$, with $\langle\Delta x\rangle = \langle\Delta y\rangle = 0$ and $\langle\left(\Delta x\right)^2\rangle = \langle\left(\Delta y\right)^2\rangle = \sigma_\mathrm{h}^2$.\cite{note1} 

This simulation has been used successfully for a couple of years in teaching contexts. More recently, we realized it could be adapted to show disorder-suppressed diffusion. The strategy was to relax the constraint $\langle\Delta x\rangle = \langle\Delta y\rangle = 0$ and, instead, segment the square domain into $m\times m$ smaller subdomains of size $1/m\times 1/m$. Each subdomain was assigned nonzero values of $\langle\Delta x\rangle$ and $\langle\Delta y\rangle$, drawn from a normal distribution: $\mathcal{N}(0,\sigma_\mathrm{d}^2)$, with $\sigma_\mathrm{d}$ a dimensionless disorder strength parameter. Physically, each subdomain represents a region with a fixed, random local drift vector that biases particle motion in one direction. This random drift field breaks the isotropy of diffusion, biasing trajectories differently in each subdomain and leading to local particle trapping.  Inevitably, some regions exhibit negative divergence and act as attractors, preferentially collecting particles, while regions of positive divergence tend to expel particles. Although this model neglects the wave interference central to AL, it provides an intuitive picture of how spatial disorder can suppress diffusion by funneling trajectories into confined regions.  

\subsection{Simulation results}
\begin{figure*}
    \centering
    \begin{tabular}{lcr}
        (a)~\includegraphics[height=0.75\columnwidth]{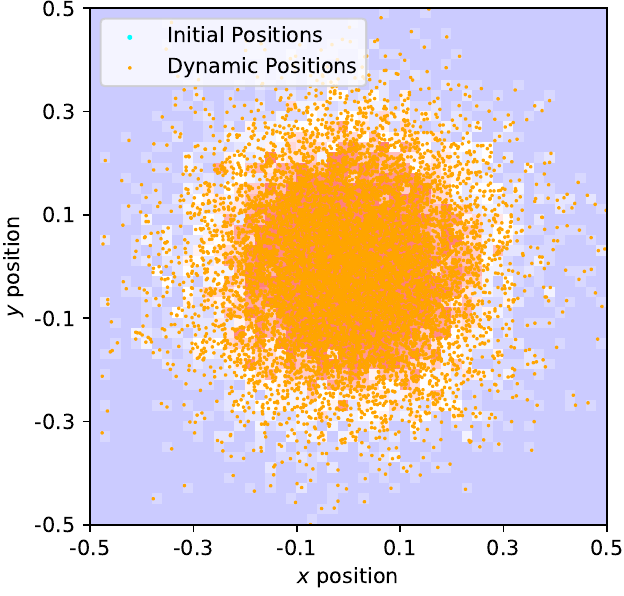} & ~\qquad~ & (b)~\includegraphics[height=0.75\columnwidth]{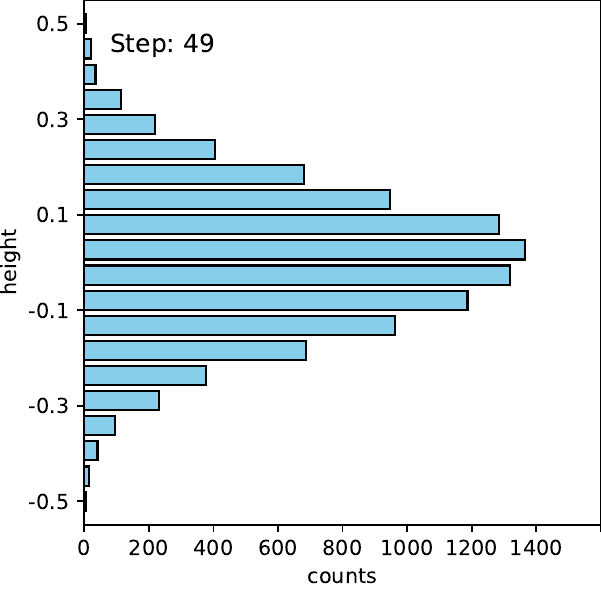}  \\
        ~ & ~\\
        (c)~\includegraphics[height=0.75\columnwidth]{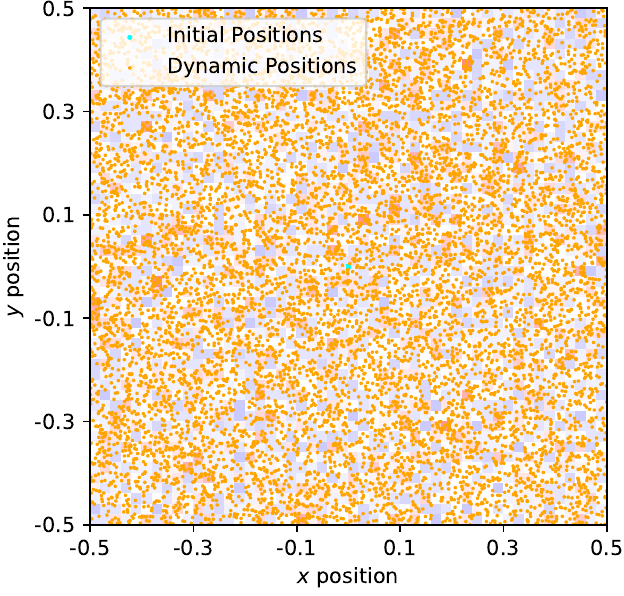} & ~\qquad~ & (d)~\includegraphics[height=0.75\columnwidth]{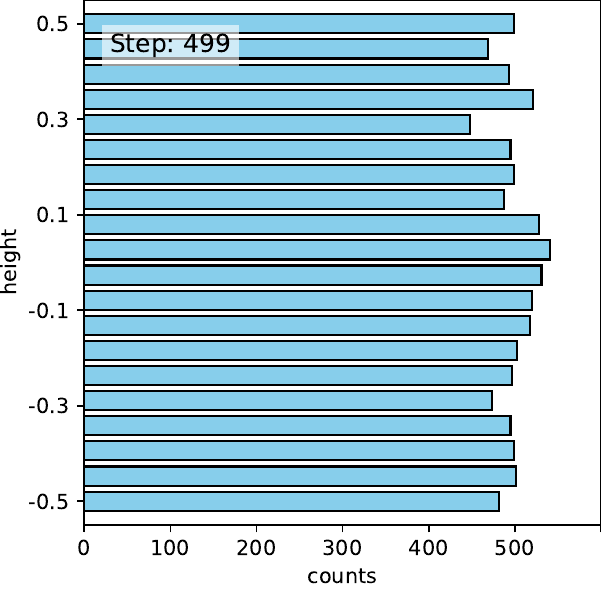}  \\
        ~ & ~\\
        (e)~\includegraphics[height=0.75\columnwidth]{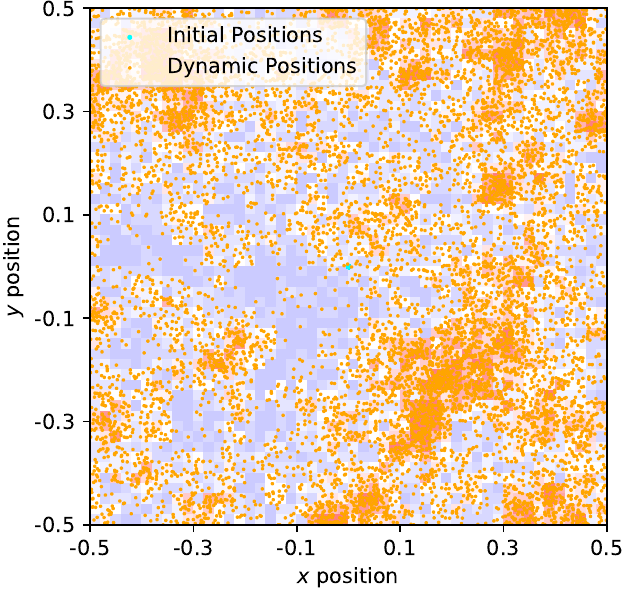} & ~\qquad~ & (f)~\includegraphics[height=0.75\columnwidth]{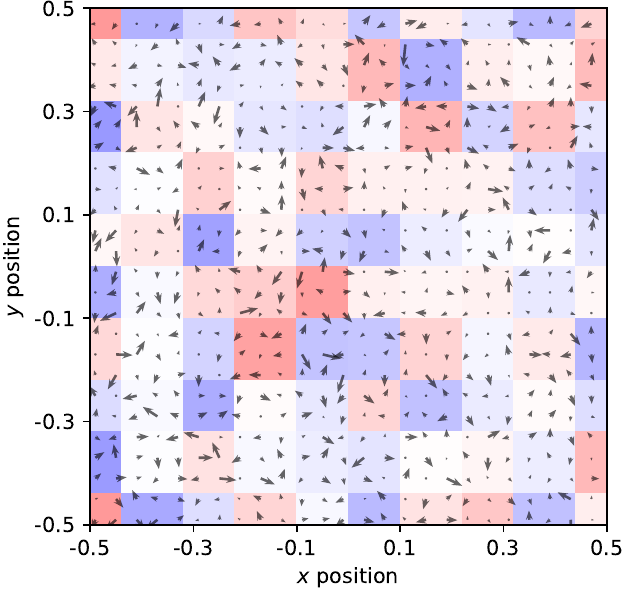} 
    \end{tabular}
    \caption{Random walk simulation results.  (a, b) Particle positions and vertical distribution after $49$ steps with $\sigma_\mathrm{d} = 0$. (c, d) The same simulation shown in (a, b) after 499 steps, showing uniform filling of the domain due to periodic boundaries.  (e) Particle density after 999 steps with $\sigma_\mathrm{d}/\sigma_\mathrm{tot} = 0.60$ showing localization.  (f) Drift vector field and divergence corresponding to (e).  Blue indicates negative divergence (sinks), red positive divergence (sources), with intensity proportional to magnitude.}
    \label{fig:Fig1}
\end{figure*}

The simulations presented in this paper were implemented in Python, with all source code made accessible via a public GitHub repository.\cite{Bobowski:2025}  Figure~\ref{fig:Fig1}(a)-(d) shows the diffusion of particles through a homogeneous region ($\sigma_\mathrm{d} = 0$). 
$10^4$ non-interacting point particles were placed at $(x, y) = (0, 0)$ and allowed to execute independent 2D random walks with step sizes drawn from a normal distribution.  Figure~\ref{fig:Fig1}(a) shows the particle positions after $49$ steps and (b) shows their vertical distribution along a line passing through the origin.  As expected, the particles follow a Gaussian distribution centered on their starting positions with $\langle x\rangle = \langle y\rangle \approx 0$ maintained throughout the simulation.  Initially, the spread of particles, characterized by $\langle\left(\Delta x\right)^2\rangle$ and $\langle\left(\Delta y\right)^2\rangle$, grew linearly with the number of steps which was reflected in a linear growth of the variance $\sigma^2$ of the distribution shown in Fig.~\ref{fig:Fig1}(b).  This linear growth was cutoff once particles reached the boundaries of the domain at $(x, y) = \pm 0.5$.  We implemented periodic boundary conditions such that, for example, a particle that reaches the right boundary reappears at left: 
\begin{equation}
(x, y) = (\pm 0.5 \pm \delta) \;\;\longrightarrow\;\; (x, y) = (\mp 0.5 \pm \delta),
\end{equation}
where $\delta\gtrsim 0$. Therefore, as shown in Fig.~\ref{fig:Fig1}(c) and (d), after a sufficient number of steps, the particles uniformly fill the space and their distribution becomes flat.

\begin{figure*}
    \centering
    \begin{tabular}{lcr}
        (a)~\includegraphics[height = 0.825\columnwidth]{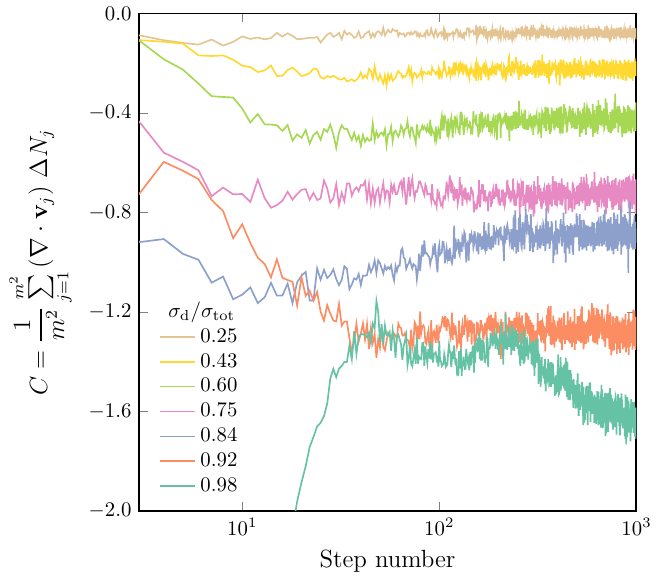}  & ~\qquad~ & (b)~\includegraphics[height = 0.825\columnwidth]{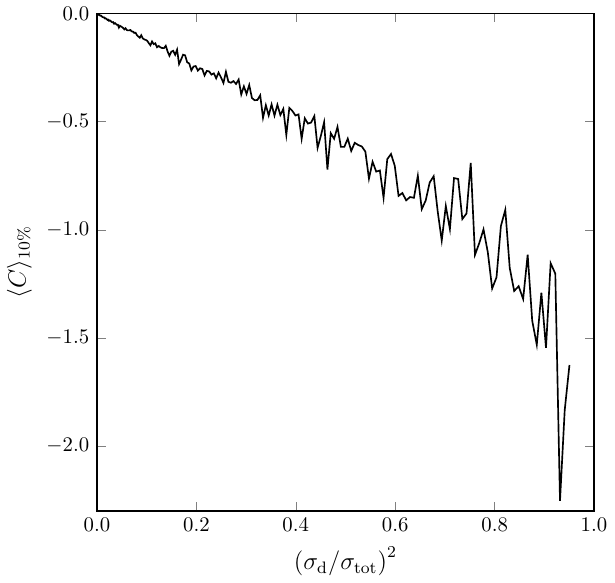}  
    \end{tabular}
    \caption{(a) Correlation score versus step number for a range of $\sigma_\mathrm{d}/\sigma_\mathrm{tot}$ values spanning very weak to strong scattering.  The $\sigma_\mathrm{d}/\sigma_\mathrm{tot} = 0.60$ corresponds to the case shown in Figs.~\ref{fig:Fig1}(e) and (f).  (b) The average of the correlation score from the final \SI{10}{\percent} of iterations as a function of $\left(\sigma_\mathrm{d}/\sigma_\mathrm{tot}\right)^2$.}
    \label{fig:Fig2}
\end{figure*}

Having verified the expected homogeneous diffusion, we next explored nonzero values of the disorder \mbox{strength $\sigma_\mathrm{d}$}.  To ensure simulations using different disorder strengths could be compared directly, we enforced the constraint $\sigma_\mathrm{h}^2 + \sigma_\mathrm{d}^2 = \sigma_\mathrm{tot}^2$, where $\sigma_\mathrm{tot} = 0.02$ was held constant so that the total variance of step sizes is fixed, ensuring that observed localization effects arise only from disorder and not from differences in overall step length for $0 <\sigma_\mathrm{d}/\sigma_\mathrm{tot} < 1$.

Figure~\ref{fig:Fig1}(e) shows the distribution of particles after 999 steps for $\sigma_\mathrm{d}/\sigma_\mathrm{tot} = 0.60$.  It is immediately clear that, even after double the number of steps shown in Fig.~\ref{fig:Fig1}(c), the particles in the disordered medium preferentially cluster in regions that localize the particles.  As will be shown shortly, this localization sets in after only 200 to 300 iterations with minimal change in the particle distribution afterwards.  The same GitHub repository that hosts the source code, also has animated GIFs of the particle distribution evolution for both the homogeneous (diffusive) and disordered (localized) cases.\cite{Bobowski:2025}   Figure~\ref{fig:Fig1}(f) is a map of the drift vector field $\mathbf{v}_j$ that led to the particle distribution in (e).  The square domain of area 1 was subdivided into $50\times 50$ subdomains with each assigned a random local drift vector.  For clarity, we show only half of the drift vectors using black arrows.  The colored squares in (f) show the average divergence of the drift vector field averaged over a $5\times 5$ set of subdomains.  Blue and red represent negative (sink) and positive (source) divergences while the darkness indicates the strength of the divergence.  A comparison of Figs.~\ref{fig:Fig1}(e) and (f) shows a tendency for particles to localize in regions of negative divergence while avoiding positive regions.  To make this connection quantitative, we next introduce a simple metric that correlates the equilibrium particle density with the divergence of the underlying drift field.

For each $m\times m$ subdomain $j$, we calculated the divergence of the drift vector field, $\nabla\cdot \mathbf{v}_j$, and the excess particle count relative to a uniform distribution, $\Delta N_j = N_j - N/m^2$. We then defined a pedagogically-motivated correlation score for the overall domain as:
\begin{equation}
C = \frac{1}{m^2}\sum_{j=1}^{m^2} 
    \bigl(\nabla \cdot \mathbf{v}_j\bigr)\,\Delta N_j .
\end{equation}
With this definition, $C < 0$ indicates that excess particles are preferentially found in regions of negative divergence (sinks), consistent with localization. By contrast, $C \approx 0$ corresponds to homogeneous diffusion.  Figure~\ref{fig:Fig2}(a) shows how the correlation score evolves with step number for a subset of the $\sigma_\mathrm{d}/\sigma_\mathrm{tot}$ values that we simulated.  As expected, we find strong anticorrelation between the local drift gradient and excess particle count as the disorder strength parameter $\sigma_\mathrm{d}$ is increased.  The simulation shown in Figs.~\ref{fig:Fig1}(e) and (f) used $\sigma_\mathrm{d}/\sigma_\mathrm{tot} = 0.6$ and corresponds to the green curve in Fig.~\ref{fig:Fig2}(a).  For this case, $C$ has reached its equilibrium value by about $200$ steps indicating that the particles have localized.  

Next, we investigated the dependence of the equilibrium values of $C$ on the disorder strength.  We did this by averaging $C$ over the last \SI{10}{\percent} of the $10^3$ steps for each value of $\sigma_\mathrm{d}/\sigma_\mathrm{tot}$ simulated, a quantity that we denote $\langle C\rangle_{10\%}$.  Figure~\ref{fig:Fig2}(b) shows that $\langle C\rangle_{10\%}$ varies linearly with  $\left(\sigma_\mathrm{d}/\sigma_\mathrm{tot}\right)^2$ from homogeneous diffusion ($\sigma_\mathrm{d}/\sigma_\mathrm{tot} = 0$) to strong disorder ($\sigma_\mathrm{d}/\sigma_\mathrm{tot}\to 1$).  This quadratic dependence is a natural consequence of how the correlation score was defined.  First, the magnitude of $\mathbf{v}_j$ and, therefore, the value of $\left\vert\nabla\cdot \mathbf{v}_j\right\vert$ is proportional to $\sigma_\mathrm{d}$.  Second, the deviation from a homogeneous distribution of particles $\left\vert \Delta N_j \right\vert$, is itself proportional to the strength of the particle sinks and sources $\left(\left\vert\nabla\cdot \mathbf{v}_j\right\vert\right)$.  Therefore, $C$, constructed from a product of the drift gradient and the particle number deviation, depends quadratically on $\sigma_\mathrm{d}$ which sets the degree of disorder. 

\begin{figure*}
    \centering
    \includegraphics[width=1.75\columnwidth]{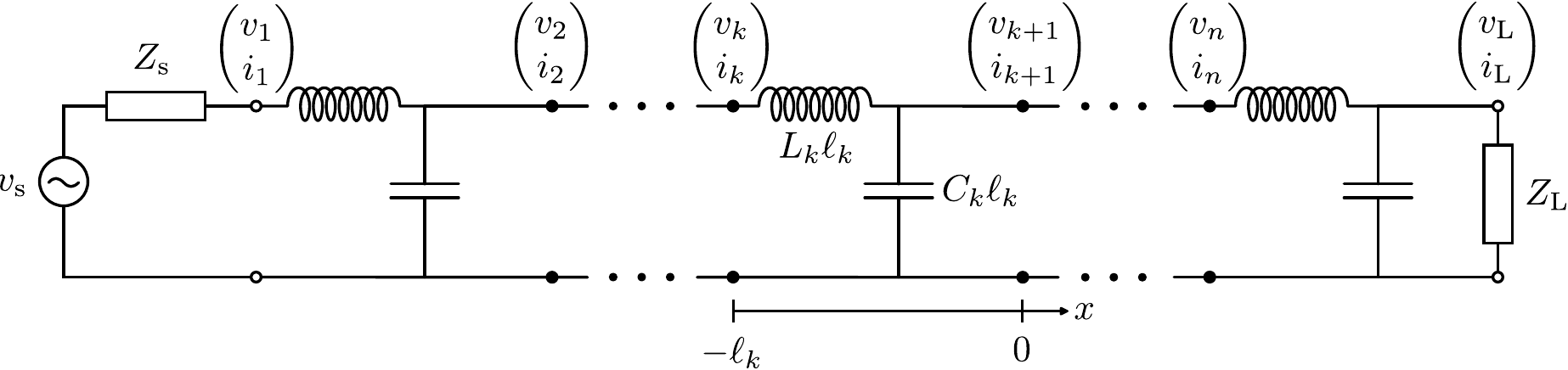}
    \caption{Lumped-element circuit model of a lossless transmission line segmented into $n$ sections of length $\ell_k$. Each section is represented by a series inductance $L_k\ell_k$ and a shunt capacitance $C_k\ell_k$, supporting currents $i_k$ and voltages $v_k$. The line is driven by a source $v_\mathrm{s}$ with source impedance $Z_\mathrm{s}$ and terminated with load impedance $Z_\mathrm{L}$. This model forms the basis for the transfer-matrix formalism used to analyze localization in disordered transmission lines.}
    \label{fig:Fig3}
\end{figure*}

The point-particle model developed in this section provides an intuitive picture of how spatial disorder can suppress diffusion by funneling trajectories into localized regions. However, it neglects the essential ingredient of Anderson localization: wave interference. In both quantum and classical wave systems, confinement arises not only from biased motion but from the destructive interference of multiply scattered waves. To capture this physics, we now turn to a 1D transmission line model.

\section{Disordered Transmission Lines and 1D Anderson Localization}
Several factors motivated us to adopt a transmission line (TL) framework. First, TLs are already covered in the lecture component of our third-year experimental physics course, so students are familiar with the concepts of characteristic impedance, propagation constants, and reflection coefficients. Second, even simulations of homogeneous TLs reveal rich physics, providing opportunities to connect with other parts of the undergraduate curriculum. Finally, while our focus here is on lossless TLs, the model can be readily extended to include conductor and dielectric losses, allowing exploration of how dissipation modifies localization phenomena.

The standard lumped-element circuit model of a lossless TL driven by a source $v_\mathrm{s}$ with output impedance $Z_\mathrm{s}$ and terminated with load impedance $Z_\mathrm{L}$ is shown in Fig.~\ref{fig:Fig3}.  The transmission line is segmented into $n$ sections and the goal is to calculate the voltage and current at the inputs and outputs of each segment.  $L_k$ and $C_k$ are the per-unit-length inductance and capacitance of segment $k$ of length $\ell_k$.  It is worth pointing out that the series inductance carries current $i_k$ which is associated with the magnetic fields in the TL at that segment position. Likewise, voltage $v_k$ appears across the shunt capacitance and is associated with the TL electric fields.  This association clarifies how the lumped-element TL captures the wave dynamics required to model AL.

Our approach builds directly on earlier analyses of transmission line transients. In previous work, we developed analytical solutions for lossless lines using Laplace-transform methods,\cite{Koyano:2023} and later extended the framework to include conductor and dielectric losses, non-ideal pulse shapes, and oscilloscope input impedances.\cite{Bobowski:2021, Bobowski:2023} These studies established that relatively simple TL models, when combined with frequency- or time-domain analysis, can capture a rich variety of transient behaviors observed in the laboratory. Here, we repurpose the same underlying circuit model to explore how randomness in the per-unit-length inductance and capacitance produces multiple scattering, destructive interference, and ultimately Anderson localization. This continuity allows us to connect experimental TL physics familiar from the advanced undergraduate lab directly to simulations that visualize localization in the time domain.

\subsection{Transfer Matrix Formalism}\label{sec:transfer}
Analysis of an arbitrary segment in isolation in the $\ell \to 0$ limit leads to the well-known telegrapher's equations:\cite{Haus:1989, Pozar:2012}
\begin{align}
    \frac{\partial v\left(x, t\right)}{\partial x} &= - L\frac{\partial i\left(x, t\right)}{\partial t},\label{eq:tele_V}\\
    \frac{\partial i\left(x, t\right)}{\partial x} &= - C\frac{\partial v\left(x, t\right)}{\partial t}.\label{eq:tele_I}
\end{align}
The time dependence can be removed if one assumes harmonic time dependencies $v(x,t)=V(x)e^{-j\omega t}$ and $i(x,t)=I(x)e^{-j\omega t}$ which leads to the wave equations for the position-dependent voltage and current {\it amplitudes}:
\begin{align}
    \frac{\partial^2 V(x)}{\partial x^2} &= -\beta^2V(x),\\
    \frac{\partial^2 I(x)}{\partial x^2} &= -\beta^2I(x),
\end{align}
where $\beta = \omega\sqrt{L C}$ is the propagation constant.  It is worth emphasizing that assuming a harmonic time dependence does not restrict the class of signals under consideration. Just as a periodic function can be represented as a Fourier series of discrete harmonics, any finite-energy (square-integrable) time-domain signal can be expressed as a continuous superposition of harmonics via the Fourier transform.
The general solution to the wave equations, with the constraints imposed by Eqs.~(\ref{eq:tele_V}) and (\ref{eq:tele_I}) are: 
\begin{align}
    V(x) &= V^+ e^{-j\beta x} + V^- e^{j\beta x},\label{eq:wave_soln1}\\
    I(x) &=\frac{1}{Z_{0}}\left[V^+ e^{-j\beta x} - V^- e^{j\beta x}\right],\label{eq:wave_soln2}
\end{align}
where $V^+$ and $V^-$ are the complex amplitudes of the forward- and backward-traveling waves, respectively, and $Z_{0} = \sqrt{L/C}$ is the segment's characteristic impedance.  

Using the coordinate system for segment $k$ indicated in Fig.~\ref{fig:Fig3}, we take the left edge of the $k$-th segment to be at $x=-\ell_k$ and the right edge at $x = 0$.  Thus, the boundary values are $V_{k} \equiv V(-\ell_k)$ and $V_{k+1} = V(0)$, with analogous definitions for $I(x)$.  These boundary conditions allow us to eliminate $V^+$ and $V^-$ from Eqs.~(\ref{eq:wave_soln1}) and (\ref{eq:wave_soln2}) and relate $V_k$ and $I_k$ to $V_{k+1}$ and $I_{k+1}$ in terms of a matrix equation:
\begin{equation}
    \begin{pmatrix}
V_k\\
I_k
\end{pmatrix}
=\boldsymbol{T}_k
\begin{pmatrix}
V_{k+1}\\
I_{k+1}
\end{pmatrix},\label{eq:trans1}
\end{equation}
where:
\begin{equation}
\boldsymbol{T}_k =\begin{pmatrix}
\cos\beta_k\ell_k & jZ_{0,k}\sin\beta_k\ell_k\\
\dfrac{j}{Z_{0,k}}\sin\beta_k\ell_k & \cos\beta_k\ell_k
\end{pmatrix}
\end{equation}
is the transfer matrix for segment $k$.

The transfer matrix relation can be applied to each segment such that:
\begin{equation}
\begin{pmatrix}
V_{1}\\
I_{1}
\end{pmatrix}
= \boldsymbol{T}_\mathrm{net} 
\begin{pmatrix}
V_\mathrm{L}\\
I_\mathrm{L},
\end{pmatrix}\label{eq:trans2}
\end{equation}
where:
\begin{align}
\boldsymbol{T}_\mathrm{net} &= \boldsymbol{T}_1 \boldsymbol{T}_2\dots \boldsymbol{T}_{n-1} \boldsymbol{T}_n\label{eq:Tnet}\\
&\equiv
\begin{pmatrix}
A_n & B_n\\
C_n & D_n
\end{pmatrix}.
\end{align}
Note that multiplication proceeds from right to left, so that $\boldsymbol{T}_n$ acts first on the load boundary conditions, followed by $\boldsymbol{T}_{n-1}$ and so on.  Using this formalism, the input impedance of the TL is:
\begin{equation}
Z_\mathrm{in} = \frac{V_1}{I_1} = \frac{A_n Z_\mathrm{L} + B_n}{C_n Z_\mathrm{L} + D_n},
\end{equation}
where $Z_\mathrm{L} = V_\mathrm{L}/I_\mathrm{L}$ is the load impedance at the termnation of the final segment.  In our AL simulations, $Z_\mathrm{L}$ is arbitrary since, in the presence of strong disorder, a signal injected at the input of a long TL will localize before reaching the far end.  However, for pedagogical clarity, we assume an open-circuit termination such that $Z_\mathrm{L}\to\infty$, giving $Z_\mathrm{in} = A_n/C_n$. For a homogeneous line of identical segments, this reduces to the familiar textbook result $Z_\mathrm{in} = -j Z_0/\tan\left(\beta\ell\right)$ by virtue of the transfer-matrix group property $\boldsymbol{T}(\ell_1)\boldsymbol{T}(\ell_2) =\boldsymbol{T}(\ell_1 + \ell_2)$.\cite{Pozar:2012, Griffiths:2001}

The choice of an open-circuit termination has two advantages.  First, it leaves us with an open system since the impedance-matched source absorbs all reflected wave energy.  Therefore, any localized energy is not a consequence of the transmission line boundary conditions.  Second, the single reflection at the open circuit leads to a recognizable interference pattern in a homogeneous TL that encodes in it the physical length of the system.  This feature will be used to validate the simulation code.

The input impedance of the TL forms a voltage divider with the source impedance $Z_\mathrm{s}$ which is driven by \mbox{$v_\mathrm{s} = V_\mathrm{s} e^{-j\omega t}$} such that:\cite{Bobowski:2021, Koyano:2023}
\begin{equation}
  \begin{pmatrix}
V_{1}\\
I_{1}
\end{pmatrix} =  \frac{V_\mathrm{s}}{Z_\mathrm{s} + Z_\mathrm{in}}
  \begin{pmatrix}
Z_\mathrm{in}\\
1
\end{pmatrix}.\label{eq:V1I1}
\end{equation}
Finally, combining Eqs.~(\ref{eq:trans1}), (\ref{eq:trans2}) and (\ref{eq:V1I1}) leads to a general expression for the voltage and current amplitudes:
\begin{equation}
  \begin{pmatrix}
V_{m}\\
I_{m}
\end{pmatrix} =  \frac{V_\mathrm{s}}{Z_\mathrm{s} + Z_\mathrm{in}}
\left[\prod\limits_{k=1}^{m-1} \boldsymbol{T}_k^{-1}\right]
  \begin{pmatrix}
Z_\mathrm{in}\\
1
\end{pmatrix},
\end{equation}
where the relation \mbox{$\left(\boldsymbol{T}_1\dots \boldsymbol{T}_{m-1}\right)^{-1} = \boldsymbol{T}_{m-1}^{-1}\dots\boldsymbol{T}_1^{-1}$} has been used, and we identify $V_{n+1} \equiv V_\mathrm{L}$, $I_{n+1} \equiv I_\mathrm{L}$.  Having achieved the goal of calculating the voltage and current at all of the TL segment nodes, we next briefly outline some of the implementation details used in all of the TL simulations followed by a discussion of the key findings.

\subsection{Simulation details}
For all simulations presented in the main manuscript, the TL is excited by an incident Gaussian
voltage pulse injected at the input and propagating in the positive $x$ direction. The pulse is
centered at $f_0 = \SI{2.8}{\giga\hertz}$ with a spectral width $\sigma_f = f_0/50$.
The simulations employ the transfer-matrix formalism described in Sec.~\ref{sec:transfer} to compute
the complex frequency-domain voltage and current amplitudes $V_k(f)$ and $I_k(f)$ at each of the $n = 500$ TL segments of length $\ell_k$.  Each segment was assigned a per-unit-length capacitance $C_k$ and inductance $L_k$ which sets the segment's propagation constant and characteristic impedance.  For each $V_k(f)$ and $I_k(f)$, we tracked $N = 2^{20}$ discrete frequencies that uniformly span $-10 f_0 \le f \le 10 f_0$.

In the simulations, the frequency-domain Gaussian pulse incident from the source was defined with the normalization of a probability density:
\begin{equation}
    V_\mathrm{s}(f) = \frac{V_0}{\sigma_f \sqrt{2\pi}}\exp\left[-\frac{1}{2}\left(\frac{f - f_0}{\sigma_f}\right)^2\right]e^{-i 2\pi f t_0},\label{eq:Vs}
\end{equation}
such that $\int V_\mathrm{s}(f)\, df = V_0$.  Consequently, both $V_{\mathrm{s}}(f)$ and $V_k(f)$ have units of volts per hertz (\si{\volt\per\hertz}), while the corresponding time-domain signals $v_k(t)$, obtained by inverse Fourier transform, have units of volts.

The $e^{-i2\pi f t_0}$ factor in Eq.~(\ref{eq:Vs}) is used to shift the Gaussian pulse in the time domain such that it is centered on $t = t_0$.  We set $t_0 = 1 / \sigma_f$ such that the width of the time-domain pulse is $\sigma_t = 1/\sigma_\omega = t_0 / (2\pi)$, ensuring that essentially all of the pulse energy is pushed to $t>0$.  Additional implementation details, including data structures and numerical optimization strategies, are available in the publicly-accessible code repository.\cite{Bobowski:2025}

\subsection{Simulation results: Homogeneous TLs}\label{sec:homogeneous}
To validate the simulation code, we begin by considering a uniform transmission line (TL) in which
$C_k = \mu_C \equiv (v_0 Z_0)^{-1}$ and $L_k = \mu_L \equiv Z_0 / v_0$ for all segments $k$.
The characteristic impedance and propagation speeds were taken to be $Z_0 = \SI{50}{\ohm}$ and $v_0 = 0.7c$, respectively, where $c$ is the vacuum speed of light.  The TL was discretized into $n=500$ segments, each of length
$\ell_k \equiv \ell = \SI{15}{\centi\meter}$.

\begin{figure*}
    \centering
    \begin{tabular}{cc}
        (a)\includegraphics[height = 0.56\columnwidth]{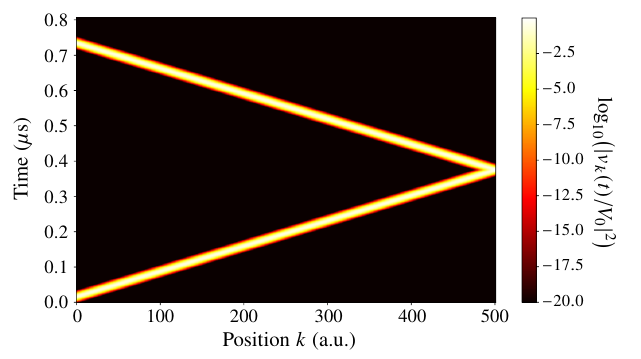} & (b)\includegraphics[height = 0.56\columnwidth]{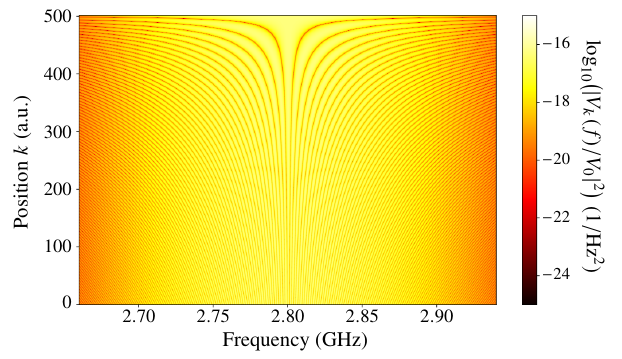} \\
        (c)\includegraphics[height = 0.56\columnwidth]{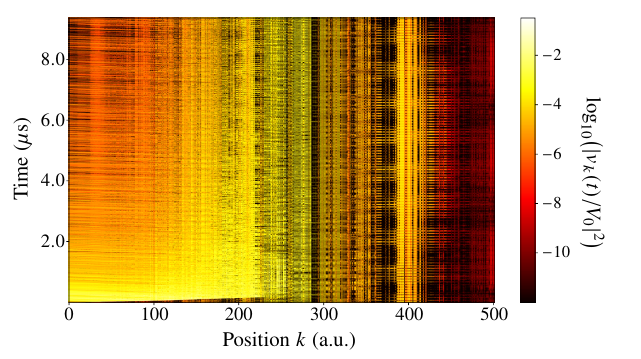} & (d)\includegraphics[height = 0.56\columnwidth]{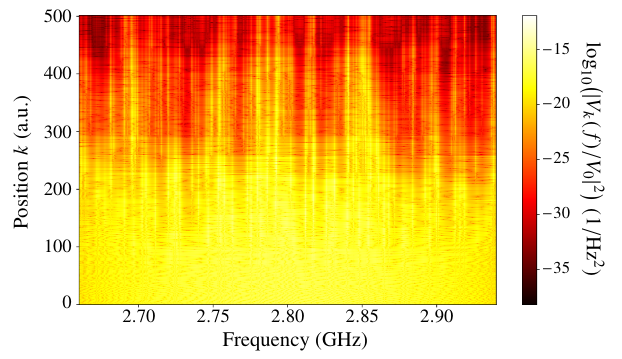} \\
        (e)\includegraphics[height = 0.56\columnwidth]{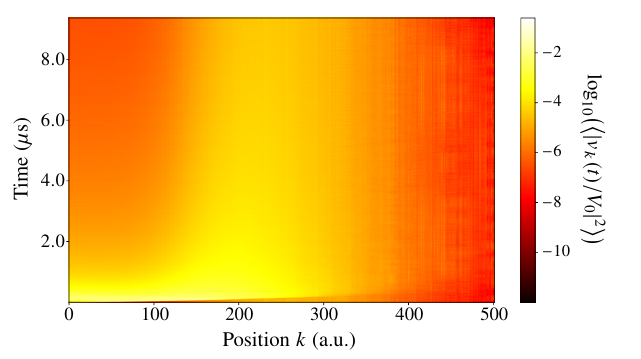} & (f)\includegraphics[height = 0.56\columnwidth]{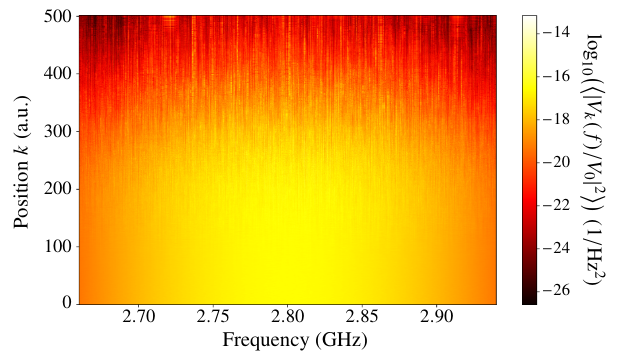} \\
    \end{tabular}
    \caption{Simulated $\left\vert v_k(t) / V_0\right\vert^2$ in the time domain and $\left\vert V_k(f) / V_0\right\vert^2$ in the frequency domain as a function of segment index $0\le k\le 500$.  In all cases, a Gaussian pulse centered on $f_0 = \SI{2.8}{\giga\hertz}$ with width $\sigma_f = f_0/50$ was incident on the TL with open-circuit termination.  (a) Ballistic propagation of a Gaussian pulse through a homogeneous TL of total length $n\ell$. (b) The frequency-domain diffraction pattern due to the ballistic pulse and open-circuit termination as a function of position in the TL. (c), (d) Single realizations of $\left\vert v_k(t) / V_0\right\vert^2$ and $\left\vert V_k(f) / V_0\right\vert^2$ for a disordered TL with $\sigma_C = \mu_C/2$, $\sigma_L = \mu_L/2$, and mean free path $\ell_\mathrm{mfp} = \SI{15}{\centi\meter}$.  The disorder is turned on slowly over a length scale of $k_\mathrm{on} = 100$ segments.  (e), (f) Ensemble averages of \si{500} realizations of $\left\vert v_k(t) / V_0\right\vert^2$ and $\left\vert V_k(f) / V_0\right\vert^2$ for the disordered TL.}
    \label{fig:Fig4}
\end{figure*}

For a homogeneous and lossless TL in the time domain, the incident Gaussian pulse is expected to:\\[-1em]
\begin{itemize}
    \setlength{\itemsep}{2pt}
    \setlength{\parskip}{0pt}
    \setlength{\parsep}{0pt}
    \item propagate ballistically through the TL along $+x$ without dispersion
    \item reflect from the open-circuit termination with reflection coefficient $\Gamma = 1$ at $t = n\ell / v_0$
    \item propagate along $-x$ and return to the impedance-matched source at $t = 2n\ell / v_0$, where it is perfectly absorbed 
\end{itemize}
Thus, for any value of $k$, the time-domain response consists of a pair of
identical Gaussian pulses separated by a time delay $\tau_d = 2\ell (n - k) / v_0$.  There is a direct analogy to a pair of optical apertures separated in space by a distance $d$.  The Fourier transform of the spatially-separated apertures leads to the familiar double-slit interference pattern in which the spacing of bright fringes is inversely proportional to $d$.  Likewise, the frequency-domain response of the TL is expected to reveal an analogous interference pattern, in which the spacing of the ``bright spots'' (large $|V_k(f)|^2$) is inversely proportional to $\tau_d$ and, hence, increases with increasing $k$.

Figures~\ref{fig:Fig4}(a) and (b) confirm all of the expected behaviors of the homogeneous TL.  In Fig.~\ref{fig:Fig4}(a), linear fits to the peak values of $\left\vert v_k(t) / V_0\right\vert^2$ confirm a propagation velocity of $\pm 0.7 c$ in agreement with the model parameters. Additionally, the measured temporal and spatial widths of the pulse at fixed position and fixed time, respectively, are consistent with the expected Gaussian envelope determined by the input bandwidth.   In Fig.~\ref{fig:Fig4}(b), the frequency-domain interference fringes exhibit the expected position-dependent spacing, scaling linearly with the inverse round-trip delay $\tau_d^{-1}\propto \left(n - k\right)^{-1}$, providing an additional consistency check of the model.

The Supplemental Materials are used to explore the time- and frequency-domain dynamics of Gaussian pulses in TLs. First, the time-domain energy distribution for the cases of homogeneous and strongly-disordered TLs are directly compared.  Second, the structure of the interference fringes shown in Fig.~\ref{fig:Fig4} is explored for a set of fixed $k$ values in order to better highlight the double-slit analogy.  Finally, we briefly consider a homogeneous TL with conductor losses and its effect on the frequency-domain response.     

Having used the homogeneous TL to validate the simulation code, we now turn our focus to disordered TLs and Anderson localization. 

\subsection{Simulation results: Disordered TLs}\label{sec:disordered}

To simulate a disordered TL, each segment $k$ was assigned a unique value of
$C_k$ and $L_k$ drawn from normal distributions with the same means $\mu_C$
and $\mu_L$ used for the homogeneous TL. The strength of disorder was controlled
via the segment-dependent standard deviations $\sigma_C$ and $\sigma_L$.
To ensure that a substantial fraction of the incident energy penetrates a
minimum depth into the TL, the disorder was turned on adiabatically:
\begin{align}
\sigma_C &= a \mu_C \left(1 - e^{-k/k_\mathrm{on}}\right)^2,\\    
\sigma_L &= a \mu_L \left(1 - e^{-k/k_\mathrm{on}}\right)^2.
\end{align}
Here, $0 \le a \le 1$ is a disorder strength parameter and $k_\mathrm{on}$ is the
disorder onset parameter. All disordered TL simulations presented here used
$k_\mathrm{on} = 100$, corresponding to \SI{20}{\percent} of the TL’s total length.

In addition, the length $\ell_k$ of each segment was randomized by drawing values
from an exponential distribution with mean $\ell_\mathrm{mfp}$. In this way,
$\ell_k$ represents the distance traveled by electromagnetic waves between
scattering events, with a mean free path of
$\ell_\mathrm{mfp} = \SI{15}{\centi\meter}$. Although randomizing $\ell_k$ is not
required to simulate Anderson localization, it more closely approximates
physically realizable disordered one-dimensional systems.

Figures~\ref{fig:Fig4}(c) and (d) show single realizations of
$\left| v_k(t) / V_0 \right|^2$ and $\left| V_k(f) / V_0 \right|^2$,
respectively, for a disorder strength of $a = 1/2$.
At early times ($t < \SI{1}{\micro\second}$),
Fig.~\ref{fig:Fig4}(c) shows that most of the energy is concentrated
in the region $k < k_\mathrm{on}$.
This behavior results from the adiabatic buildup of disorder,
which suppresses strong backscattering of the incident pulse
immediately after it is launched.

At later times, the bulk of the energy is concentrated between
$100 < k < 250$, with an additional localized contribution near
$k \approx 400$ for this particular realization.
The time axis extends beyond
$\SI{9}{\micro\second} \gg n\ell_\mathrm{mfp}/v_0$,
the characteristic time required for a ballistic pulse to reach
the end of the TL.
Moreover, the spatial distribution of energy is nearly static
for times $t > \SI{2}{\micro\second}$.
Together, these observations provide the first clear indications
of Anderson localization.

It is also noteworthy that essentially no energy reaches the end
of the TL, implying that for sufficiently long transmission lines
with strong disorder, the choice of load termination has a
negligible effect on the simulation results.

Figure~\ref{fig:Fig4}(d) is proportional to the spectral distribution of energy
in the TL integrated over the observation time. At all values of $k$, the
spectral energy is concentrated within a narrow band of frequencies centered
near $f_0$, corresponding to the bandwidth of the incident pulse. For any fixed
frequency within this band, the spectral energy is exponentially suppressed, by
up to \num{20} orders of magnitude, as $k$ increases from zero to \num{500}.
Ultimately, the exponential tails of an ensemble average of many realizations of
$\left| V_k(f) / V_0 \right|^2$ are used to quantify the Anderson localization
length scale $\xi$.

Ensemble averages of \num{500} realizations of
$\left| v_k(t) / V_0 \right|^2$ and $\left| V_k(f) / V_0 \right|^2$ are shown in
Figs.~\ref{fig:Fig4}(e) and (f), respectively, for the same disorder strength
($a = 1/2$) used in Figs.~\ref{fig:Fig4}(c) and (d). These ensemble averages
confirm the interpretation of the single realizations while simultaneously
smoothing sharp features, thereby enabling a meaningful quantitative analysis
of the simulated results.

\begin{figure}
    \centering
    \includegraphics[height = 0.85\columnwidth]{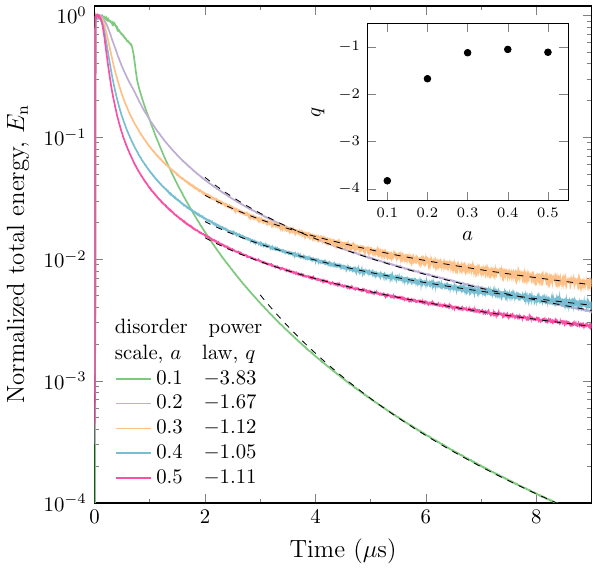} \\
    \caption{Total energy, normalized to its maximum value, as a function of time for disordered TLs with the disordered strength parameter set to  $a = 0.1$, \num{0.2}, \num{0.3}, \num{0.4} and \num{0.5}.  The dashed lines are power law fits ($A t^q$) to the asymptotic tails.  As the disorder strength is increased, more energy is localized in the TL at long times marking a clear crossover into the Anderson localization regime.  Inset: The power $q$ as a function of disorder strength $a$.}
    \label{fig:Fig5}
\end{figure}

The normalized total energy in the TL at time $t$ can be obtained from
time-domain data via:
\begin{align}
E(t) &\propto \sum_{k=0}^{N}\left\langle \left|v_k(t)\right|^2 \right\rangle,\\
E_\mathrm{n}(t) &= \frac{E(t)}{\max[E(t)]},
\end{align}
where the denominator denotes the maximum value of $E(t)$ over the observation
time. The magenta line in Fig.~\ref{fig:Fig5} shows $E_\mathrm{n}(t)$ for the
$a = 0.5$ data of Fig.~\ref{fig:Fig4}(e). This analysis was repeated for
$a = 0.1$ to \num{0.4} in steps of \num{0.1}. In each case, ensemble averages
of \num{500} realizations were used to extract the normalized energy.
Figure~\ref{fig:Fig5} displays $E_\mathrm{n}(t)$ versus $t$ for all five values
of $a$ used in the simulations.  
In all cases, the normalized energy quickly reaches its peak before dropping off.  The initial rapid decline in energy is due to strong backscattering once $k \gtrsim k_\mathrm{on}$.  These backward-traveling waves readily reach the source and exit the TL due to the quasi-homogeneous section of the TL for $k < k_\mathrm{on}$.  

Later times probe waves that penetrate deeper into the TL.  For $t\gtrsim \SI{3}{\micro\second}$, the rate of energy escape drops dramatically, particularly for the largest disorder strengths.  The dashed lines are power-law ($At^q$) fits to the large-$t$ tails of $E_\mathrm{n}(t)$ and demonstrate a clear deviation from exponential decay, consistent with power-law energy leakage in open 1D localized systems. The best-fit values of $q$ are plotted as a function of $a$ in the inset of the Fig.~\ref{fig:Fig5} and clearly exhibit a crossover to an Anderson localization regime, with $q$ asymptotically approaching $\approx -1.1$, indicating increasingly suppressed transport and long-lived energy trapping with increasing disorder.  Although the data do show Anderson localization, notice that the total confined energy is never constant.  With one end of the TL connected to an impedance-matched source, there is always a small, but finite, probability of energy eventually scattering back to the source.
\begin{figure}
    \centering
    \includegraphics[height = 0.85\columnwidth]{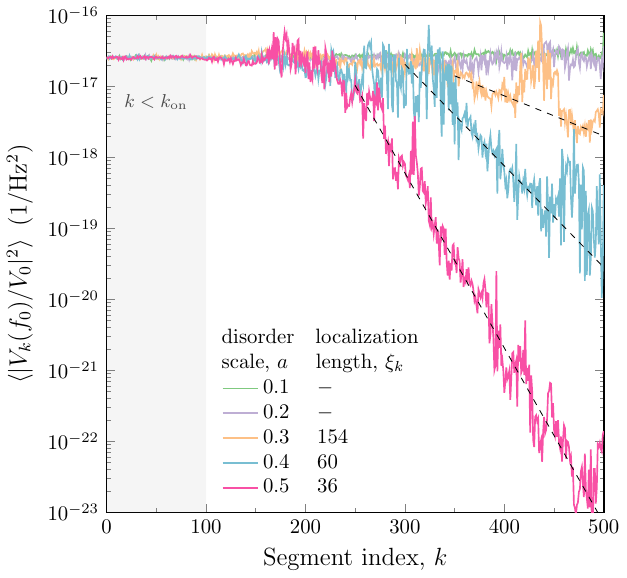} \\
    \caption{$\langle\left\vert V_k(f_0) / V_0\right\vert^2\rangle$ as a function of $k$ for $f_0=\SI{2.8}{\giga\hertz}$.  The shaded region spanning $0\le k\le 100$ marks the region over which the disorder is adiabatically activated.  The dashed lines are exponential fits to the large-$k$ tails ($B\exp(-2 k / \xi_k)$) where $\xi_k$ is the localization length.  As in Fig.~\ref{fig:Fig5}, there is a clear crossover into the Anderson localization regime as the disorder strength is increased.}
    \label{fig:Fig6}
\end{figure}
This behavior contrasts with closed systems, where localized states can retain energy indefinitely.\cite{Allen:1998,Lee:1985}

For the same scattering strengths examined in Fig.~\ref{fig:Fig5}, Fig.~\ref{fig:Fig6} shows the ensemble-averaged spectral energy at $f_0 = \SI{2.8}{\giga\hertz}$ over a \SI{53}{\kilo\hertz} bandwidth as a function of $k$. 
These data are obtained from vertical slices through the spectral heatmaps at $f_0$ and should be interpreted as showing the spatial distribution of spectral energy remaining in the TL long after the incident pulse has been launched.  The magenta line ($a = 0.5$) of Fig.~\ref{fig:Fig6} corresponds to a vertical slice through Fig.~\ref{fig:Fig4}(f).  

For $k < k_\mathrm{on}$, the scattering is very weak for all values of $a$ and all of the spectral energies share a common baseline.  This collapse onto a single baseline reflects the quasi-homogeneous nature of the TL for 
$k < k_\mathrm{on}$, where the incident wave propagates with negligible scattering regardless of disorder strength.  On the other hand, the $k > k_\mathrm{on}$ behavior is sensitive to the scattering strength.  For $a\le 0.2$, $\langle\left\vert V_k(f_0) / V_0\right\vert^2\rangle$ is independent of $k$ indicating that the spectral energy is approximately uniform after ensemble averaging.  For these scattering strengths, the incident pulse disperses weakly as it travels along the TL, allowing it to largely preserve its Gaussian shape, albeit with an ever-increasing width. Consequently, the pulse reflects from the open end and subsequently returns to the source.  After the ensemble average, any spectral energy that does remain in the TL due to weak scattering is uniformly distributed along the line.

For $a > 0.2$, the onset of Anderson localization is marked by the appearance of an extended region over which the spectral energy decays exponentially with increasing $k$.  A characteristic localization length $\xi$ is typically defined via an exponential decay of wave intensity, $\vert V(x)\vert^2 \propto e^{-2x/\xi}$.  For our simulation data, the tails of $\langle\left\vert V_k(f_0) / V_0\right\vert^2\rangle$ were fit to $Be^{-2 k / \xi_k}$ in order to extract a dimensionless localization parameter $\xi_k$ defined in terms of the TL segment number.  To convert to a length, $\xi_k$ can be multiplied by the assumed mean free path $\ell_\mathrm{mfp} = \SI{15}{\centi\meter}$.  The dashed lines in Fig.~\ref{fig:Fig6} show the exponential fits and the best-fit values for $\xi_k$ are given in the plot legend.  For the $a=0.3$ case, the peak that appears in the exponential tail from, $k\approx 400$ to $450$ was excluded from the fit.

Although the TL is terminated by an open circuit, the absence of significant spectral energy at large $k$ indicates that reflections from the termination do not play a dominant role in shaping the exponential tails used to extract $\xi_k$.  Taken together, Figs.~\ref{fig:Fig5} and \ref{fig:Fig6} demonstrate that the crossover from diffusive transport to Anderson localization manifests consistently in both the long-time decay of the total energy and the exponential suppression of the spectral energy with position.

\subsection{Discussion}\label{sec:discuss}
The \hyperref[sec:intro]{Introduction} described quantum scattering amplitudes for electron transport, the problem that motivated Anderson's original localization theory. Here, we discuss an alternative coupled-normal-mode viewpoint that the segmented-TL simulation highlights particularly well. In the simulation, each TL segment is assigned randomized values of characteristic impedance $Z_k$, propagation speed $v_k$, and length $\ell_k$. The combination $\beta_k = \omega/v_k = \omega\sqrt{L_k C_k}$ defines the segment's propagation constant (wavenumber), where $L_k$ and $C_k$ are the per-unit-length inductance and capacitance, respectively, of the underlying discrete circuit model of the TL.

The coupling between segments is set by impedance mismatch. At first glance, the coupling is limited to nearest-neighbor interactions via a local reflection coefficient $\Gamma_k = (Z_{\mathrm{eff},k} - Z_k) / (Z_{\mathrm{eff},k} + Z_k)$. However, the linear chain of all downstream segments determines the effective impedance $Z_{\mathrm{eff},k}$ terminating segment $k$. In this way, all segments are recursively coupled, with the effective coupling strength weakening with separation distance. This mutual coupling between segments is naturally captured by the transfer-matrix formalism presented in Sec.~\ref{sec:transfer}.

Each segment has separate reflection coefficients for forward- and backward-traveling waves, denoted $\Gamma_k^+$ and $\Gamma_k^-$. In the idealized limit of strong reflections, if $\operatorname{sgn}(\Gamma_k^+ \Gamma_k^-) = +1$, the boundary conditions admit local standing-wave--like modes with $\beta_k \approx m\pi/\ell_k$, while $\operatorname{sgn}(\Gamma_k^+ \Gamma_k^-) = -1$ gives $\beta_k \approx (m-1/2)\pi/\ell_k$, where $m$ is a positive integer. Consequently, depending on the local boundary conditions, the disordered TL can be viewed as a linear array of resonators with randomized normal-mode frequencies and coupling strengths.

In extended systems of many coupled resonators, energy transport is not governed by simple nearest-neighbor hopping between individual segments, but by interference between many hybridized normal modes that, in the absence of disorder, extend along the entire length of the TL. As disorder increases, mode hybridization becomes increasingly local, ultimately confining energy to a finite subset of nearly-matched resonators that efficiently exchange among one another. This confinement of energy into local clusters of closely-matched segments suppresses transport and provides an intuitive understanding of the exponential decay of spectral energy along the TL observed in Fig.~\ref{fig:Fig6}.

\section{Summary}\label{sec:summary}
We have presented a pair of relatively simple numerical simulations to explore the physics of Anderson localization from a pedagogical perspective, without relying on graduate-level field-theoretic calculations.

We began with a noninteracting point-particle random walk to simulate diffusive transport. In the absence of disorder, a collection of particles placed at a single point diffuses outward, creating a Gaussian density profile with a constant mean and a variance that grows linearly in time. Disorder was implemented by dividing the square domain of the simulation into subdomains that preferentially scatter particles in randomized directions. Clusters of subdomains with inward-pointing scattering gradients act as sinks that suppress diffusive transport by localizing particles. Although the point-particle simulations do not capture the wave-interference effects that are central to Anderson localization, they build intuition for how disorder, in the absence of dissipation, can suppress transport and lead to localization-like behavior.

The second simulation used a segmented transmission line to model Anderson localization in a one-dimensional wave system. Disorder was introduced by randomizing the characteristic impedance, propagation speed, and length of each segment. Impedance mismatches between segments lead to scattering at segment boundaries, with wave-interference effects tracked using a transfer-matrix formalism. A frequency-domain Gaussian pulse was launched, and the complex voltage and current amplitudes were calculated at each segment node. Ensemble averages over many disorder realizations were used to generate time- and frequency-domain energy distributions for various disorder strengths.

The time-domain results were used to determine the total energy in the transmission line as a function of time. As disorder was increased, the rate of energy escape from the open 1D system was strongly suppressed at times long after the incident pulse was launched. The persistence of confined energy for times much longer than the ballistic time-of-flight provides clear evidence of a crossover from diffusive transport to Anderson localization in an open system. In the frequency domain, the spectral energy was found to be exponentially suppressed along the transmission line, with an increasingly shorter localization length as the disorder strength was increased.

The transmission-line framework also lends itself to straightforward extensions. In particular, conductor and dielectric losses can be included through complex propagation constants, with physically-motivated frequency dependencies, allowing students to explore how dissipation competes with interference and modifies the apparent localization length. Such extensions provide a natural bridge between idealized Anderson localization and transport in real, lossy systems.




\begin{acknowledgments}
We gratefully acknowledge the computational support provided by the Digital Research Alliance of Canada.
\end{acknowledgments}

\clearpage{\thispagestyle{empty}\cleardoublepage}



\section*{Supplementary material}

dielectric losses can be included through complex propagation constants, with physically-motivated frequency dependencies, allowing students to explore how dissipation competes with interference and modifies the apparent localization length. Such extensions provide a natural bridge between idealized Anderson localization and transport in real, lossy systems.

# Simulating diffusion and disorder-induced localization: supplemental material

Jake S. Bobowski[*]
*Department of Physics, University of British Columbia,
3333 University Way, Kelowna, British Columbia, V1V 1V7, Canada*

(Dated: January 4, 2026)

This document provides supplemental material to the main manuscript entitled *Simulating diffusion and disorder-induced localization in random walks and transmission lines*. The material presented here focuses on the transmission line simulations. First, we compare and contrast the time-domain propagation of an incident Gaussian pulse in homogeneous and disordered transmission lines. Second, the analogy between frequency-domain interference arising from a pair of pulses separated in time and the familiar double-slit interference pattern from optics is discussed in greater detail. Finally, we examine how loss can be incorporated into simulations of homogeneous transmission lines and its effect on the resulting frequency-domain interference pattern.

## S1. TIME-DOMAIN PULSE PROPAGATION

Figure 4(a) of the main manuscript shows a heatmap of $|v_k(t)/V_0|^2$ as a function of time and position for a homogeneous transmission line (TL). Figure S1(a) shows horizontal slices through the heatmap at five fixed times that span $0 < t < 2n\ell/v_0$, which is the time required for the Gaussian pulse to ballistically travel twice the length of the TL constructed from $n = 500$ identical segments. The arrows next to each pulse specify the instantaneous direction of propagation. The position $k = 500$ corresponds to the end of the TL which is terminated by an open circuit with a reflection coefficient $\Gamma = 1$. At this position, the forward and backward traveling pulses momentarily overlap which doubles $|v_k(t)|$ and quadruples $|v_k(t)|^2$. The Gaussian pulses in Fig. S1(a) have been normalized such that $|v_k(t)/V_0|^2 = 1$ at $t = n\ell/v_0$ or, equivalently, at $k = 500$.

In the frequency domain, the width of the Gaussian pulse was set to $\sigma_f = f_0/50$, such that the time-domain width is $\sigma_t = (2\pi\sigma_f)^{-1} = 25/(\pi f_0)$ which corresponds to a dimensionless spatial width of $\sigma_k = \sigma_t v_0/\ell$. Using $f_0 = 2.8$ GHz, $v_0 = 0.7c$ and $\ell = 15$ cm gives an expected $\sigma_k \approx 4$ which is a very good match to the Gaussian pulses observed in Fig. S1(a).

Figure S1(b) is the complementary data for time-domain pulse propagation in a disordered transmission line with a disorder strength of $a = 1/2$. These plots were generated by taking horizontal slices through the heatmap of Fig. 4(c) in the main manuscript which corresponds to a single realization of the disordered TL simulation. The data of Fig. S1(b) have the same normalization as those in (a). Note, however, that, for clarity, the vertical scale in (b) is logarithmic and the span of $k$ has been reduced.

For the two earliest times the pulse predominantly preserves its Gaussian profile since, for these times, $k < k_{\mathrm{on}}$ and the pulse is propagating through the quasi-homogeneous section of the TL. The vertical dashed line in Fig. S1(b) marks the position of $k_{\mathrm{on}}$. The leading edge of the $t = 103.7$ ns pulse is just beyond $k_{\mathrm{on}}$. There are remnants of the Gaussian profile at the front edge,

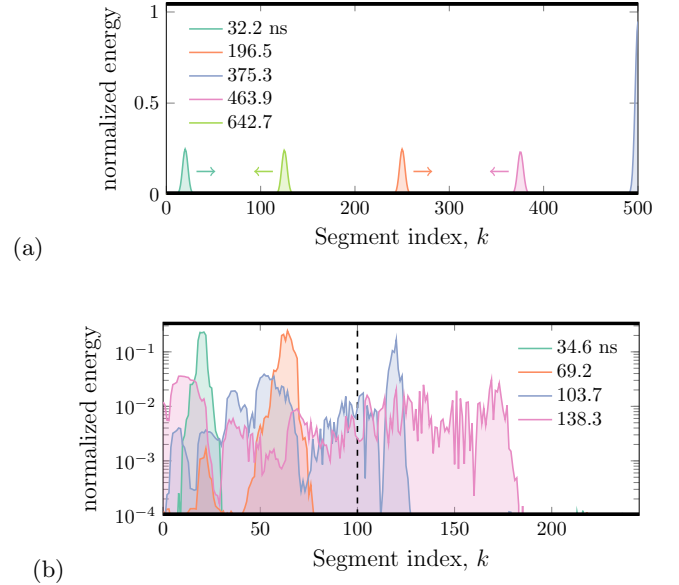

FIG. S1. (a) Snapshots of the ballistic propagation of a Gaussian pulse through a homogeneous TL. At the open-end ($k = 500$), where $\Gamma = 1$, there is a superposition of the forward and backward traveling pulses. (b) Propagation of an incident Gaussian pulse through a disordered TL with $\sigma_C = \mu_C/2$, $\sigma_L = \mu_L/2$, and $\ell_{\mathrm{mfp}} = 15$ cm.

but strong backscattering is marked by the weak, but extended, trailing tail in the $k < k_{\mathrm{on}}$ region of the TL. By $t \gtrsim 138.3$ ns, the Gaussian profile is completely washed out due to strong scattering.

The code repository contains animated GIFs showing the propagation of an incident Gaussian pulse in both homogeneous and disordered TLs.[S1]

## S2. DOUBLE-SLIT ANALOGY

Everything required to explicitly calculate the frequency-domain interference pattern of a pair of identical pulses separated in time by $\tau_d$ was presented in Sec. IV A of the main manuscript. The strategy is

to treat the homogeneous transmission line terminated by an open circuit as a single segment of total length $\ell_{\text{tot}} = n\ell$. In this case, Eq. (14) of the main manuscript gives:

$$Z_{\text{in}} = \frac{A_n}{C_n} = -jZ_0 \cot(\beta \ell_{\text{tot}}), \qquad (S1)$$

where the propagation constant $\beta = \omega/v_0$. Equation (15) of the main manuscript then gives the voltage amplitude at the TL input as $V_1(\omega) = V_s(\omega)/(Z_s/Z_{\text{in}} + 1)$, where $V_s(\omega)$ is the frequency-domain pulse incident from the source and we take the source impedance $Z_s = Z_0$ to be matched to the characteristic impedance of the TL. Combining these results leads to:

$$V_1(\omega) = \frac{V_s(\omega)}{1 + j\tan(\beta \ell_{\text{tot}})} \qquad (S2)$$

$$= \frac{V_s(\omega)}{2}\left(1 + e^{-2j\beta\ell_{\text{tot}}}\right), \qquad (S3)$$

where the identity $1 + j\tan x = (\cos x + j\sin x)/\cos x$ has been used. Taking the absolute square to obtain a quantity proportional to intensity, gives:

$$|V_1(\omega)|^2 = \frac{|V_s(\omega)|^2}{2}\left[1 + \cos(2\beta\ell_{\text{tot}})\right] \qquad (S4)$$

$$= |V_s(\omega)|^2 \cos^2\left(\frac{\omega\tau_d}{2}\right), \qquad (S5)$$

where, in the last step, we have used $\beta\ell_{\text{tot}} = \omega\ell_{\text{tot}}/v_0 = \omega\tau_d/2$. Equation (S5) has precisely the same form as the familiar double-slit interference pattern produced by a pair of identical optical apertures separated in space by a distance $d$.

The $\cos^2(\omega\tau_d/2)$ factor determines the positions of the interference maxima and minima in frequency, while the $|V_s(\omega)|^2$ prefactor determines the shape of the intensity envelope. Just as the geometry of the optical apertures sets the shape of the intensity envelope, the profile of the frequency-domain pulse likewise sets the envelope of the spectral energy. Our simulations used a Gaussian pulse which leads to a Gaussian envelope. Had we chosen a rectangular pulse in the time domain, the frequency-domain envelope would have a $\text{sinc}^2(\omega\tau_a)$ dependence, where $\tau_a$ is the temporal width of the pulse (analogous to a rectangular aperture of width $a$).

Figure S2 shows $|V_k(f)|^2$ normalized to $|V_k(f_0)|^2$ for several values of $k$ along a homogeneous TL. The values of $k$ used match the pulse positions chosen for Fig. S1(a) and share a common color scheme. These plots were generated by taking horizontal slices through Fig. 4(b) of the main manuscript.

The Gaussian envelope, as expected, is independent of position $k$ and has a width $\sigma_f = f_0/50 = 56\,\text{MHz}$. The fringe spacing, on the other hand, increases with increasing $k$. As discussed in the main manuscript, the effective temporal spacing between the incident and reflected pulses decreases as $k$ increases: $\tau_d = 2\ell(n-k)/v_0$. This

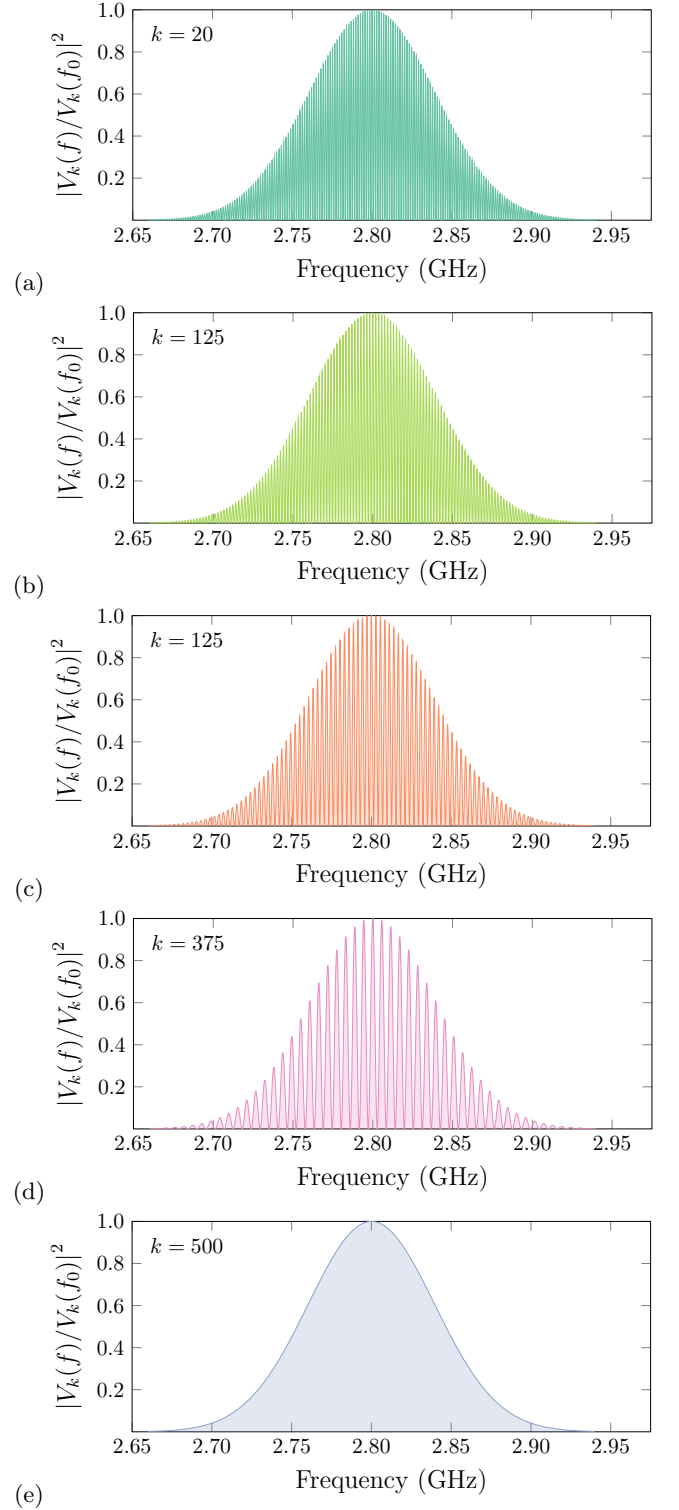

FIG. S2. Snapshots of $|V_k(f)/V_k(f_0)|^2$ as a function of frequency for a ballistic Gaussian pulse propagating through a homogeneous TL. The positions and color sequence match those used in Fig. S1. The density of the interference fringes is inversely proportional to the distance from the open-circuit termination at $k = 500$.

result can be used to generalize Eq. (S5) for arbitrary $k$:

$$|V_k(\omega)|^2 = |V_s(\omega)|^2 \cos^2\left[\frac{\omega\ell(n-k)}{v_0}\right], \quad \text{(S6)}$$

such that constructive interference occurs when $\omega\ell(n-k)/v_0 = m\pi$, with $m$ integer. Alternatively, the fringe spacing in frequency is given by:

$$\Delta f = \frac{v_0}{2\ell(n-k)}. \quad \text{(S7)}$$

Figure S2 clearly shows an increasing $\Delta f$ as $k$ increases. Furthermore, when $k = n$, $\Delta f \to \infty$ such that the interference pattern vanishes and only the Gaussian envelope remains.

## S3. LOSSES IN A HOMOGENEOUS TL

Here, we briefly discuss the inclusion of loss in the transmission-line (TL) circuit model and then demonstrate one effect of conductor loss using our segmented-TL simulation. Everything in the main manuscript, as well as all material presented in this supplemental document up to this point, has assumed a lossless TL composed solely of ideal series inductors and shunt capacitors.

In a real TL, conductor loss is modeled by a per-unit-length resistance placed in series with the inductors. Dielectric loss, by contrast, is modeled by a shunt conductance placed in parallel with the capacitors, which admits a leakage current to ground.

In principle, the per-unit-length resistance and conductance may be parameterized arbitrarily. For example, in real coaxial cables operated at RF, the cross-sectional area through which current flows is restricted by the electromagnetic skin depth $\delta \propto \omega^{-1/2}$, leading to a resistance that scales as $R \propto \omega^{1/2}$.[S2,S3]

For simplicity, we assume a constant $R > 0$ and set $G = 0$ in our TL simulation. Since we employ a relatively narrowband Gaussian pulse in the frequency domain, any weak frequency dependence of $R$ would lead only to minor quantitative corrections to the results presented here.

Loss is incorporated into the simulation by allowing the inductance to become complex according to $L \to L(1-jr)$, where $r$ is a dimensionless loss parameter analogous to the imaginary component of the complex relative permeability of the core material inside a solenoid. The advantage of this approach is that no other modifications to the code are required. Both the characteristic impedance $\sqrt{L/C}$ and propagation constant $\omega\sqrt{LC}$ then naturally become complex, leading to phase shifts and attenuation.

To preserve a frequency-independent series resistance, we take $r = r_0\omega_0/\omega$, such that the inductive impedance becomes:

$$Z_L = j\omega L + r_0\omega_0 L, \quad \text{(S8)}$$

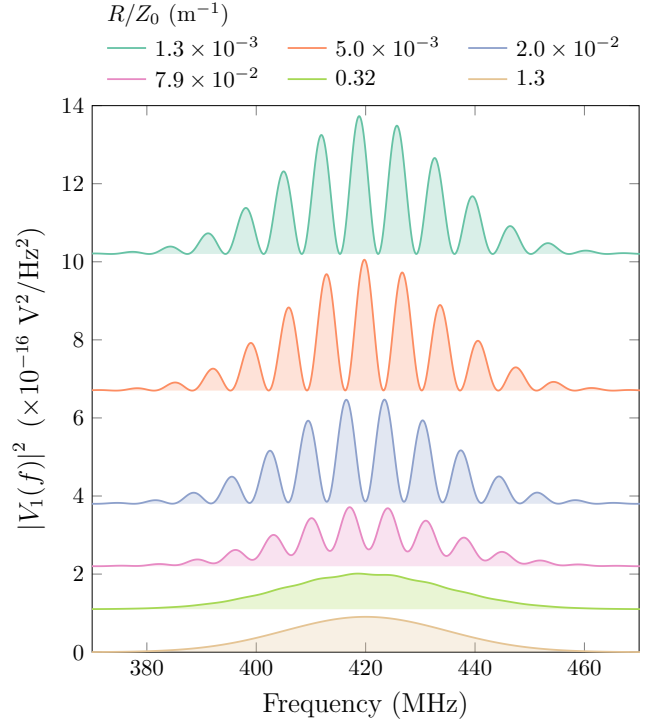

FIG. S3. Spectral energy density as a function of frequency at the TL input with varying levels of conductor loss incorporated into the TL model. For clarity, subsequent curves have been offset vertically while maintaining a consistent vertical scale for each dataset. As conductor losses are increased: (1) the amplitude of the diffraction envelope decreases, (2) the depth of the diffraction minima decreases, and (3) the phase of the interference pattern shifts.

and hence $R \equiv r_0\omega_0 L$. We repeated the frequency-domain simulations of the homogeneous TL interference pattern discussed in Sec. S2 after adding various amounts of conductor loss, measured in fractions of $Z_0$.

For the simulations with conductor loss, we used $f_0 = 420\,\text{MHz}$, $\sigma_f = f_0/20$, and a TL made of 100 segments of mean length $\ell = 15\,\text{cm}$. These choices ensure that the fringe spacing $\Delta f = 7\,\text{MHz}$ at the TL input is easily resolved in plots of $|V_1(f)|^2$ as a function of frequency. Figure S3 shows the simulation results for six logarithmically-spaced values of conductor loss spanning $1.3 \times 10^{-3} \le R/Z_0 \le 1.3\,\text{m}^{-1}$. For clarity, subsequent curves have been offset vertically while maintaining a consistent scale.

The lowest values of $R/Z_0 = 1.3 \times 10^{-3}\,\text{m}^{-1}$ and $5.0 \times 10^{-3}\,\text{m}^{-1}$ produce only subtle changes in $|V_1(f)|^2$, indicating that conductor loss acts as a weak perturbation in this regime. The most noticeable effect is a small phase shift in the locations of the fringe maxima and minima. Increasing $R/Z_0$ to $0.020\,\text{m}^{-1}$ enhances this phase shift and attenuates the spectral peaks. Further increases to $R/Z_0$ progressively broaden the fringe pattern, lifting the minima above zero and eventually washing out the interference structure entirely.




\begin{thebibliography}{99}
\bibitem{note1} A simple extension of this simulation is achieved by setting $\langle \Delta y\rangle$ equal to a negative constant.  This causes the particles to drift downwards in order to mimic settling due to a constant gravitational potential.  After a sufficient number of iterations, the vertical distribution of particles approaches an exponential distribution.
\end{thebibliography}
\end{document}